\documentclass[prl,aps]{revtex4}
\usepackage{graphicx,bm}
\begin{document}

\title{Spin and Current Variations in Josephson Junctions}

\author{A. Shnirman$^{1}$, Z. Nussinov$^{2}$, Jian-Xin Zhu$^{2}$,
A. V. Balatsky$^{2}$, and Yu. Makhlin$^{1,3}$}

\affiliation{$^{1}$ Institut f\"ur Theoretische
Festk\"orperphysik, Universit\"at Karlsruhe, D-76128 Karlsruhe,
Germany} \affiliation{$^{2}$ Theoretical Division, Los Alamos
National Laboratory, Los Alamos, New Mexico 87545, USA}
\affiliation{$^{3}$ Landau Institute for Theoretical Physics,
Kosygin st. 2, 117940 Moscow, Russia}

\begin{abstract}
We study the dynamics of a single spin embedded in the tunneling
barrier between two superconductors. As a consequence of pair
correlations in the superconducting state, the spin displays
rich and unusual dynamics. To properly describe the time evolution
of the spin we derive the effective Keldysh action for the spin. 
The superconducting correlations lead to an
effective spin action, which is non-local in time, leading to
unconventional precession. We further illustrate how the
current is modulated by this novel spin dynamics.
\end{abstract}
\maketitle

\section{Introduction}

The analysis of spins embedded in Josephson junctions has had a
long and rich history.
Early on, Kulik~\cite{Kulik_JETP66} argued that spin flip processes in
tunnel barriers reduce the critical Josephson current
as compared to the Ambegaokar-Baratoff
limit~\cite{Ambegaokar_Baratoff}. More than a decade later,
Bulaevskii et al.~\cite{Bulaevskii_pi_junction} conjectured that
$\pi$-junctions may be formed if spin flip processes dominate.
The competition between the Kondo effect and the superconductivity was
elucidated in \cite{Glazman_Matveev_SC_Kondo}. Transport
properties formed the central core of these and many other
pioneering works, while spin
dynamics was relegated to a relatively trivial secondary role.
In the current article, we report on new non-stationary spin dynamics
and illustrate that the spin is affected
by the Josephson current. As a consequence of
the Josephson current, spins exhibit novel non-planar
precessions while subject to the external magnetic field.
A spin in a magnetic field exhibits circular Larmor precession 
about the direction of the field.
As we report here, when the spin
is further embedded between two superconducting leads, new out-of-plane
longitudinal motion, much alike that displayed by a mechanical
top, will arise. We term this new effect the
{\em Josephson nutation}. We further outline how transport is,
in turn, modulated by this rather unusual spin
dynamics. Our predictions are within experimental
reach, and we propose a detection scheme.

\section{The system}

\begin{figure}
\centerline{\includegraphics[width=0.25\columnwidth]{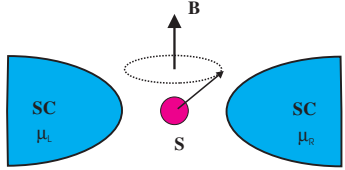}}
\caption{Magnetic spin coupled to two superconducting leads.}
\label{FIG:SETUP}
\end{figure}

The system under consideration is illustrated in
Fig.~\ref{FIG:SETUP}. It consists of two identical
ideal s-wave superconducting
leads coupled each to a single spin; the
entire system is further subject to a weak external
magnetic field. In Fig.(\ref{FIG:SETUP}), $\mu_{L,R}$ denote
the chemical potentials of the left and right leads,
${\bf{B}}$ is a weak external magnetic field
along the z-axis,
and ${\bf {S}} = (S_{x}, S_{y},S_{z})$ is
the operator of the localized spin.
The Hamiltonian of the system
reads
\begin{eqnarray}
&&{\cal H}={\cal H}_0+{\cal H}_T, \ \ \
{\cal H}_0={\cal H}_{L}+{\cal H}_{R}-\mu B_{z}S_z, \\
&&{\cal H}_T=\sum_{\mathbf{k},\mathbf{p},\alpha,\alpha'} e^{i\phi/2}\;
c_{R\mathbf{k}\alpha}^{\dagger}\left[T_0\delta_
{\alpha\alpha'}+T_1\,\mbox{\boldmath$\sigma$}_{\alpha\alpha'}\cdot{\bf
S}\,\right]\,c_{L\mathbf{p}\alpha'}+h.c.,
\end{eqnarray}
where ${\cal H}_{L}$ and ${\cal H}_{R}$ are the Hamiltonians in
the left and right superconducting leads, while $c^{\dagger}_{ik\alpha}$
($c_{ik\alpha}$) creates (annihilates) an electron in the lead a
in the state $\mathbf{k}$ with spin $\alpha$ in the right/left
lead for $i=L/R$ respectively. The vector
$\mbox{\boldmath$\sigma$}$ represents the three Pauli matrices
and $\mu$ is the magnetic moment of the spin.  When a spin is embedded
in the tunneling barrier, the conduction electron tunneling matrix becomes,
not too surprisingly,
spin-dependent \cite{Balatsky_Martin,Balatsky_Manassen}
$\hat{T}=[T_0 \hat 1 + T_1 \mathbf{S} \cdot
\hat{\bm{\sigma}}].$
Here $T_0$ is a spin-independent tunneling matrix element and
$T_1$ is a spin-dependent matrix element originating from the
direct exchange coupling $J$ of the conduction electron to the
localized spin ${\bf S}$. We take both tunneling matrix elements
to be momentum independent.
This is not a crucial assumption and is merely introduced to
simplify notations. Typically, from the expansion of the work
function for tunneling, $\frac{T_1}{T_0} \sim J/U$, where $U$ is the height of 
a spin-independent tunneling barrier~\cite{Zhu_Balatsky}.
A weak external magnetic field $B_{z} \sim 100 Gauss$ will not influence
the superconductors and we may
ignore its effect on the leads. In what follows, we abbreviate
$\mu B_z$ by $B$. The operator $e^{i\phi/2}$ is the (single electron) number
operator. When the junction is linked to an external environment, the
coupling between the junction and the environment induces fluctuation 
of the superconducting phase $(\phi)$.

\section{The Effective Action}

Josephson junctions are necessarily
embedded into external electrical circuits.
This implies that the dynamics
will explicitly depend on the superconducting phase
$\phi$. The evolution operator is given by the real-time path integral
\begin{equation}
Z = \int D\phi D{\bf{S}} \; \exp\left[i \mathcal{S_{\rm
circuit}}(\phi)+i\mathcal{S_{\rm spin}}({\bf{S}})+
i\mathcal{S_{\rm tunnel}(\phi,{\bf{S}})}\right] \ .
\end{equation}
The effective action $\mathcal{S_{\rm tunnel}}$ describes the junction itself.
We generalize the formerly known effective tunneling action for a spin-less
junction~\cite{AES_82,Larkin_Ovchinnikov_83,ESA_84} to the spin-dependent
arena to obtain
\begin{eqnarray}
\label{EQ:S_EFF} \mathcal{S_{\rm tunnel}}  &=& - 2 \oint_K dt\oint_K
dt'\, \alpha(t,t')\,\left[T_0^2 + T_1^2 {\bf S}(t)\cdot{\bf
S}(t')\right]\, \cos\left[{\frac{\phi(t) -\phi(t')}{2}}\right]
\nonumber \\ &-& 2 \oint_K dt\oint_K dt'\,
\beta(t,t')\,\left[T_0^2 - T_1^2 {\bf S}(t)\cdot{\bf
S}(t')\right]\, \cos\left[{\frac{\phi(t) +\phi(t')}{2}}\right] \ ,
\end{eqnarray}
where $i\alpha(t,t')\equiv G(t,t')G(t',t)$, $i\beta(t,t')\equiv
F(t,t')F^{\dag}(t,t')$ and the Green functions
\begin{eqnarray}
&&G(t,t')\equiv -i\sum_\mathbf{k}\,\langle T_{K}
c_{\mathbf{k}\sigma}^{\phantom{\dagger}}(t)
c_{\mathbf{k}\sigma}^{\dagger}(t') \rangle\\
&&F(t,t')\equiv
-i\sum_\mathbf{k}\,\langle T_{K}
c_{\mathbf{k}\uparrow}(t)c_{\mathbf{-k}\downarrow}(t') \rangle\\
&&F^{\dag}(t,t') \equiv -i \sum_\mathbf{k} \langle T_{K}\,
c^{\dag}_{\mathbf{k}\uparrow}(t)\, c^{\dag}_{-\mathbf{k}\downarrow}(t') \rangle
\ .
\end{eqnarray}

In Eq.~(\ref{EQ:S_EFF}) $\oint_K$ denotes integration
along the Keldysh contour. We
now express the spin action on Keldysh
contour in the basis of coherent states
\begin{equation}
\mathcal{S}_{\rm spin} = \oint_{K} dt {\bf{B}} \cdot {\bf{S}} +
\mathcal{S}_{WZNW} \ .
\label{wzwn}
\end{equation}
Here, $S$ denotes the magnitude
of the spin ${\bf{S}}$. The second,
Wess-Zumino-Novikov-Witten (WZNW), term in Eq.(\ref{wzwn})
depicts the Berry phase
accumulated by the spin as a result of motion of the spin on a
sphere of radius $S$~\cite{Fradkin_book,Sachdev_book}. Explicitly,
\begin{equation}
\label{EQ:WZNW}
\mathcal{S}_{WZNW}=\frac{1}{S^2} \int_0^1d\tau \oint_{K} dt\,[{\bf
S}(t,\tau) \cdot (\partial_{\tau}{\bf S}(t,\tau)\nonumber
\\ \times \partial_t{\bf
S}(t,\tau))] \ .
\end{equation}
The additional integral over $\tau$ allows us
to express the action in a local form. At $\tau =0$ the
spin is set along the $z$ direction at all times, ${\bf
S}(t,0)=\mbox{const}$; at $\tau = 1$
the spin field corresponds to the physical configurations,
${\bf S}(t,1)={\bf S}(t)$.

\section{Dynamics}

We now perform the Keldysh rotation, defining
the values of the spin and the phase variables
on the forward/backward branches of the Keldysh contour
(e.g., ${\bf S}^{u,l}$ for the upper and lower branch) and rewriting all the
expressions in terms of their average (classical component ${\bf S}$)
and difference (quantum component ${\bf l}$):
\begin{equation}
{\bf S} \equiv ({\bf S}^u + {\bf S}^l)/2, \;\; {\bf l} \equiv {\bf
S}^u - {\bf S}^l,\;\; {\bf S} \cdot {\bf l} = 0.
\label{EQ:Sfields}
\end{equation}
After the Keldysh rotation
we obtain
\begin{figure}
\includegraphics[width=0.25\columnwidth]{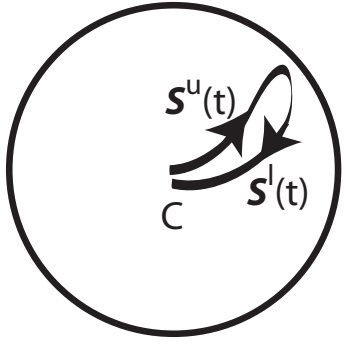}
\caption{The sphere or radius $S$ for the vectors ${\bf S}^{u,l}(t)$ is
shown. The path $C$ describes the evolution of the spin along the upper ($u$) 
and lower ($l$) branches of the Keldysh contour. To properly describe the spin 
dynamics
on this closed contour we analyze the WZNW action, see
Eq.~(\ref{EQ:WZNW}). For clarity, we draw a small piece of the
closed trajectories.} \label{FIG:WZNW}
\end{figure}
\begin{eqnarray}
\mathcal{S}_{WZNW}=\frac{1}{S^2} \int_0^1d\tau \int dt[{\bf
S}^u(t,\tau) \cdot (\partial_{\tau}{\bf S}^u(t,\tau)
\times \partial_t{\bf
S}^u(t,\tau)) -(u\rightarrow l)]\;. \label{WZW_diff}
\end{eqnarray}
The relative minus sign stems from the backward time ordering on
the return part of $C$. The individual WZNW phases for the
upper ($u$) and lower ($l$) branches are given by the areas
spanned by the trajectories ${\bf{S}}^{u,l}(t)$ on the sphere or
radius $S$ divided by the spin magnitude ($S$). The WZNW term
contains {\em odd} powers of ${\bf l}$. Insofar as the WZNW term
of Eq.(\ref{WZW_diff}) is concerned, the standard Keldysh
transformation to the two classical and quantum fields,
$\mathbf{S}$ and $\mathbf{l}$, mirrors the decomposition of the
spin in an anti-ferromagnet (AF) to the two orthogonal slow and
fast fields \footnote{ These two orthogonal AF fields represent
(i) the slowly varying staggered spin field (the antiferromagnetic
staggered moment ${\bf{m}}$ taking on the role of $\mathbf{S}$)
and (ii) the rapidly oscillating uniform spin field ${\bf l}$
(paralleling our $\mathbf{l}$). In the antiferromagnetic
correspondence, the two forward time spin trajectories at two
nearest neighbor AF sites become the two forward ($u$) and
backward ($l$) single spin trajectories of the non-equilibrium
problem. This staggered doubling correspondence is general.}. The
difference between the two individual WZNW terms in
Eq.(\ref{WZW_diff}) is the area spanned between the forward and
backward trajectories. For close
forward and backward trajectories the WZNW action on the Keldysh
loop may be expressed as
\begin{eqnarray}
\mathcal{S}_{WZNW} = \frac{1}{S^2} \int dt\, {\bf{l}} \cdot
({\bf{S}} \times \partial_{t} {\bf{S}}).
\end{eqnarray}
For the spin part of the (semi-classical) action we, then, obtain
\begin{eqnarray}
\mathcal{S}_{\rm spin} = \int dt
{\bf{B}} \cdot {\bf{l}} +  \frac{1}{S^2} \int dt\, {\bf{l}} \cdot
({\bf{S}} \times \partial_{t} {\bf{S}}).
\end{eqnarray}

Next, we perform the Keldysh rotation to the classical and quantum
components with respect to both the phase and spin variables in
the tunneling part of the effective action. Towards this end, we
introduce (with notations following Refs.~\cite{AES_82,ESA_84})
\begin{equation}
\phi \equiv (\phi^u + \phi^d)/2 \ \ , \ \ \chi \equiv \phi^u -
\phi^d \ .
\end{equation}
With these definitions in hand, the tunneling part
of the action reads
\begin{equation}
\label{EQ:S_tunnel} \mathcal{S}_{\rm tunnel} =
\mathcal{S}_{\alpha} + \mathcal{S}_{\beta} \ ,
\end{equation}
where the normal (quasi-particle) tunneling part
$\mathcal{S}_{\alpha}$ is expressed via the Green functions
$\alpha^{R}\equiv \theta(t-t')(\alpha^{>} - \alpha^{<})$ and
$\alpha^{K}(\omega) \equiv \alpha^{>} + \alpha^{<}$, where
$i\alpha^{>}(t,t') \equiv G^{>}(t,t')G^{<}(t',t)$ and
$i\alpha^{<}(t,t') \equiv G^{<}(t,t')G^{>}(t',t)$. Similarly the
Josephson-tunneling part $\mathcal{S}_{\beta}$ is expressed via
the off-diagonal Green's functions $\beta^{R}\equiv
\theta(t-t')(\beta^{>} - \beta^{<})$ and $\beta^{K}(\omega) \equiv
\beta^{>} + \beta^{<}$, where $i\beta^{>}(t,t') \equiv
F^{>}(t,t')F^{\dag >}(t,t')$ and $i\beta^{<}(t,t') \equiv
F^{<}(t,t')F^{\dag <}(t,t')$. In this paper we are
interested in the interaction between the supercurrent and the
spin. Thus we provide the expression for the Josephson part:
\begin{eqnarray}
\label{EQ:S_beta} \mathcal{S}_{\beta} &=& \int dt \int
dt^{\prime}\;4\beta^{R}(t,t^{\prime})\times
\nonumber\\
&&\left[\left\{
\left[2 T_0^2 - 2 T_1^2 {\bf{S}} (t) \cdot
{\bf{S}}(t^{\prime})\right]\sin\frac{\chi(t)}{4}\cos\frac{\chi(t')}{4}
-\frac{1}{2}\,T_1^2\; {\bf{l}}(t)\cdot
{\bf{l}}(t^{\prime}) \,\cos\frac{\chi(t)}{4}\sin\frac{\chi(t')}{4}
\right\}\sin\frac{\phi(t)+\phi(t')}{2}\right.
\nonumber \\
&&\left. +\left\{T_1^2\; {\bf{l}} (t) \cdot
{\bf{S}}(t^{\prime})\,\cos\frac{\chi(t)}{4}\cos\frac{\chi(t')}{4}
-T_1^2\; {\bf{S}}(t)\cdot
{\bf{l}}(t^{\prime}) \,\sin\frac{\chi(t)}{4}\sin\frac{\chi(t')}{4}
\right\}\cos\frac{\phi(t)+\phi(t')}{2} \right]
\nonumber\\
&+&
\int dt \int dt^{\prime}\;\beta^{K}(t,t^{\prime})\times
\nonumber\\
&&\left[\left\{
\left[4 T_0^2 - 4 T_1^2 {\bf{S}} (t) \cdot
{\bf{S}}(t^{\prime})\right]\sin\frac{\chi(t)}{4}\sin\frac{\chi(t')}{4}
+T_1^2\; {\bf{l}}(t)\cdot
{\bf{l}}(t^{\prime}) \,\cos\frac{\chi(t)}{4}\cos\frac{\chi(t')}{4}
\right\}\cos\frac{\phi(t)+\phi(t')}{2}\right.
\nonumber \\
&&\left. -\left\{2 T_1^2\; {\bf{l}} (t) \cdot
{\bf{S}}(t^{\prime})\,\cos\frac{\chi(t)}{4}\sin\frac{\chi(t')}{4}
+2 T_1^2\; {\bf{S}}(t)\cdot {\bf{l}}(t^{\prime})
\,\sin\frac{\chi(t)}{4}\cos\frac{\chi(t')}{4}
\right\}\sin\frac{\phi(t)+\phi(t')}{2} \right] \ .
\end{eqnarray}
The normal-tunneling part $\mathcal{S}_{\alpha}$ is obtained from
$\mathcal{S}_{\beta}$ by the following substitution:
$\beta^{R/K}(t,t') \rightarrow \alpha^{R/K}(t,t')$, $\phi(t')
\rightarrow -\phi(t')$, and $\chi(t') \rightarrow -\chi(t')$. The
Keldysh terms (those including $\beta^{K}$ and $\alpha^{K}$),
which normally give rise to random Langevin terms (see, e.g.,
Ref.~\cite{ESA_84}) are, in our case, suppressed at temperatures
much lower than the superconducting gap ($T \ll \Delta$), due to
the exponential suppression of the correlators
$\beta^{K}(\omega)$ and $\alpha^{K}(\omega)$ at $\omega <
\Delta$.

To obtain $\beta^{R}$ we start from the Gor'kov Green functions
\begin{equation}
F^{>}(t,t') = -i \sum_k \frac{\Delta}{2 E_k} e^{-iE_k (t-t')} \ \
\ ,\ \ \ F^{>\dag}(t,t')  = i \sum_k \frac{\Delta}{2 E_k} e^{-iE_k
(t-t')} \ ,
\end{equation}
where the quasi-particle energy
$E_k\equiv \sqrt{\Delta^2 + \epsilon_k^2}$, $\epsilon_k$
being the free-conduction-electron dispersion in the leads. Putting
all of the pieces together, we find that
\begin{equation}
\beta^{R}(t-t')=\theta(t-t') \sum_{k,p} \frac{\Delta^2}{2 E_k E_p}
\sin\left[(E_k+E_p)(t-t')\right]\ .
\end{equation}
The kernel $\beta^{R}(t-t')$ decays on (short) time scales
of order ${\cal{O}}(\hbar/\Delta)$.
Varying the full action with respect to the
quantum components $\mathbf{l}$ and $\chi$ and setting these
to zero, we obtain coupled equations of motion for
both the spin and phase:
\begin{eqnarray}
\label{Eq:Eq_motion_spin} &&\frac{d {\bm{S}(t)}}{dt} = {\bm{S}(t)}
\times {\bf{B}} + T_1^2\int dt'\,4\beta^{R}(t-t')\,{\bm{S}(t)}
\times {\bm{S}(t')}\,
\cos\frac{\phi(t)+\phi(t')}{2}\ , \\
\label{Eq:Eq_motion_phase} && \frac{\delta \mathcal{S}_{\rm
circuit}}{\delta\chi(t)}\,\vert_{\chi \rightarrow 0} = -\int
dt'\,2\beta^{R}(t-t')\,\left(T_0^2 -
T_1^2{\bm{S}(t)} \cdot {\bm{S}(t')}\right)\,
\sin\frac{\phi(t)+\phi(t')}{2}\ .
\end{eqnarray}
Note, that, if the rest of the circuit contains dissipative
elements, e.g., resistors, then $\mathcal{S}_{\rm circuit}$ will
contain the non-vanishing Keldysh components, and one should
include the corresponding Langevin terms into
Eq.~(\ref{Eq:Eq_motion_phase}). The rather complicated equations
of motion (\ref{Eq:Eq_motion_spin}) and (\ref{Eq:Eq_motion_phase})
are very general. To make headway, we now adopt a perturbative
strategy. In Eq.(\ref{Eq:Eq_motion_spin}), we first assume an
ideal voltage bias, i.e., an imposed phase $\phi(t) = \omega_{\rm
J} t$, where the ``Josephson frequency'' $\omega_{\rm J} =
2eV/\hbar$. To this lowest order, we neglect the influence of the
spin on the phase. Next, we use the separation of characteristic 
time scales to our advantage. To this end, we note that the
spin dynamics is much slower as compared to electronic processes,
i.e. $\omega_{\rm J},B \ll \Delta$. This separation of scales
allows us to set ${\bf{S}}(t^{\prime}) \simeq {\bf{S}}(t) +
(t^{\prime} - t) d{\bf{S}}/dt$ in the integrand of
Eq.(\ref{Eq:Eq_motion_spin}), wherein we obtain
\begin{equation}
\label{EQ:SPIN_EQ_MOTION} \frac{d {\bf{S}}}{dt} = \lambda {\bf{S}}
\times \frac{d {\bf{S}}}{dt}  \sin \omega_{J} t +  {\bf{S}} \times
{\bf{B}}.
\end{equation}
Here,
\begin{eqnarray} \lambda &=& -4 T_1^2 \int\limits_{0}^{\infty}
d t\cdot t\cdot \beta^{R}(t)\cdot\sin\frac{\omega_{\rm J}t}{2}
\approx -2 T_1^2 \omega_{\rm J} \int\limits_{0}^{\infty} d t\cdot
t^2\cdot \beta^{R}(t) = 2\omega_{\rm J}\sum_{k,p}\frac{\vert
\Delta\vert^{2}\vert T_{1}\vert^{2}}{E_{k}E_{p}(E_{k}+E_{p})^3}
\nonumber\\
&=& 2T_1^2 \rho^2\,\frac{\omega_{\rm J}}{\Delta}\, \int\int
\frac{d z_1 d z_2}{(\cosh z_1+\cosh z_2)^3} =
\frac{\pi^2}{8}\,T_1^2 \rho^2\,\frac{\omega_{\rm
J}}{\Delta}=\frac{g_1}{32}\,\frac{\omega_{\rm J}}{\Delta}\ ,
\label{l3}
\end{eqnarray}
with $g_1 \equiv (2\pi
T_1\rho)^2$ the spin channel conductance.
In Eq.(\ref{l3}) we employ
the separation of time scales ($\omega_{\rm J} \ll \Delta$) again.
When expressed in the spherical coordinates (in the semi-classical
limit)
${\bf S} = S( \sin \theta \cos \phi, \sin \theta \sin \phi,
\cos \theta)$,
Eq.(\ref{EQ:SPIN_EQ_MOTION})
transforms into two simple first order differential equations
\begin{eqnarray}
&&\frac{d \phi}{d t}
=-\frac{B}{1+S^2\lambda^{2}\sin^{2}(\omega_{J}t)}\ ,
\\
&&\frac{d \theta}{dt} = - S\lambda\, \frac{d \phi}{dt} \sin \theta
\sin \omega_{J}t \ .
\end{eqnarray}
\begin{figure}
\centerline{\includegraphics[width=0.25\columnwidth]{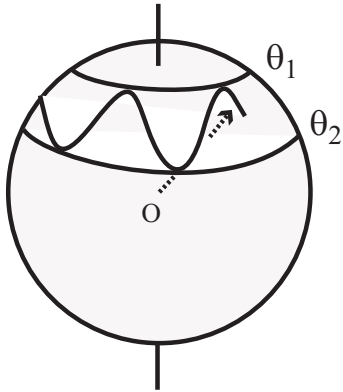}}
\caption{The resulting spin motion on the unit sphere in the
general case. As in the motion of classical spinning top, the spin
exhibits undulations along the polar direction. }
\label{FIG:RIGID}
\end{figure}
These equations can be solved exactly.
For a spin oriented at time $t=0$
at an angle $\theta_{0}$ relative
to ${\bf B}$,
\begin{eqnarray*}
\phi(t)  &=& - \frac{B}{\omega_{J}\sqrt{1+ S^{2} \lambda^{2}}}
\tan^{-1}[ \sqrt{1+ S^{2} \lambda^{2}} \tan(\omega_{J}t)], \nonumber
\\
\theta(t) &=& 2\tan^{-1}\Big( \biggl{[} \frac{(1-c\cos
(\omega_{J}t))(1+c)}{(1+c\cos(\omega_{J}t))(1-c)}
\biggr{]}^{\gamma} \tan \frac{\theta_{0}}{2} \Big)\;,
\end{eqnarray*}
with $c= S \lambda/\sqrt{1+ S^{2} \lambda^{2}}$ and $\gamma=-S
\lambda B/2\omega_{J}c$. For $S \lambda \ll 1$, $\phi \simeq -B t$
and $\theta \simeq \theta_0 - S \lambda
(B/\omega_{J})\sin\theta_0\cos\omega_J t$. Typically, whenever a
spin is subjected to a uniform magnetic field, the spin
azimuthally precesses with the Larmor frequency $\omega_{L}=B$. In
a Josephson junction, however, the spin exhibits additional polar
($\theta$) displacements. The resulting dynamics may be likened to
that of a rotating rigid top. The Josephson current leads to a
non-planar gyroscopic motion ({\em Josephson nutations}) of
the spin much like that generated by torques applied  to a
mechanical top. For small $\lambda$, we find nutations (see
Fig.~\ref{FIG:RIGID}) of amplitude $\theta_1-\theta_2 \propto
\sin\theta \cdot S\cdot\lambda \cdot\frac{B}{\omega_{\rm J}}
\propto \sin\theta \cdot S\cdot g_1 \cdot\frac{B}{\Delta}$.

The origin of the first term on the rhs of Eq.~(\ref{EQ:SPIN_EQ_MOTION}) can be 
understood as follows (this origin can be also traced in the calculations): the 
spin is subject to the electron-induced fluctuating field ${\bf h}=T_1 \Sigma 
e^{i\phi/2} c^\dagger \bm{\sigma} c + \mbox{\rm h.c.}$. The same coupling may be 
thought of as an influence of the spin on the leads, which results in a non-zero 
low-frequency contribution  $\delta\bf h$to $\bf h$. Since the response function 
of the electron liquid is isotropic but retarded, $\delta\bf h(t)$ is not 
aligned with $\bm{S}(t)$ but contains information about the values of ${\bm 
S}(t')$ at earlier times. The response function decays on a time scale $\sim 
\hbar/\Delta$, much shorter than the period of the spin precession, $\sim 1/B$. 
As a consequence, in addition to a contribution $\propto \bm {S}$ the field 
$\bm{h}$ acquires a component $\propto \dot{\bm{S}}/\Delta$, which leads to the 
first term on the rhs of Eq.~(\ref{EQ:SPIN_EQ_MOTION}).

The rhs of the second equation of motion
(\ref{Eq:Eq_motion_phase}) clearly corresponds to the Josephson
current. Indeed, in the Keldysh formalism one has $I=(2\pi/\Phi_0)
\partial \mathcal{S}/\partial \chi$ (instead of $I=(2\pi/\Phi_0)
\partial \mathcal{S}/\partial \phi$). Thus we obtain for the
Josephson current
\begin{equation}
\label{EQ:Josephson_Current}
I_{\rm J}(t) =
\frac{2\pi}{\Phi_0}\,\int dt'\,2\beta^{R}(t-t')\,\left(T_0^2 -
T_1^2{\bm{S}(t)} \cdot {\bm{S}(t')}\right)\,
\sin\frac{\phi(t)+\phi(t')}{2}\ .
\end{equation}
We start from the lowest order (local in time) adiabatic
approximation, i.e. we set $\mathbf{S}(t)=\mathbf{S}(t')$ and
$\phi(t)=\phi(t')$. This yields
\begin{equation}
\label{EQ:JC_local}
I_{\rm J}(t) = \frac{2\pi}{\Phi_0}\,E_{\rm J,0}\left(1 -
\frac{T_1^2}{T_0^2}\mathbf{S}^2\right) \sin\phi(t)\ ,
\end{equation}
where $E_{\rm J,0}\equiv 2T_0^2 \int dt
\beta^{R}(t)=\pi^2\rho^2T_0^2\Delta=(1/4) g_0 \Delta$ is the
spin-independent Josephson energy \cite{Ambegaokar_Baratoff}
($g_0$ being the conductance of the spin-independent channel).
The second term of Eq.~(\ref{EQ:JC_local}) gives the spin-related 
reduction of the Josephson critical current studied in 
Ref.~\cite{Kulik_JETP66}. 
We now evaluate the lowest-order correction to this equation
due to deviations from locality in time and spin precessions.
Expanding $\mathbf{S}(t')$ in Eq.~(\ref{EQ:Josephson_Current})
in $(t'-t)$ and using the fact that for the Larmor precession 
we have $\mathbf{S\cdot \dot S}=0$ and 
$\mathbf{S}\cdot \mathbf{\ddot S} = B^2\,(S_z^2 - \mathbf{S}^2)$,
we find a correction to the Josephson current which depends on 
$S_z^2$:
\begin{equation}
\label{EQ:JC_corrections} I_{\rm J}(t) =
\frac{2\pi}{\Phi_0}\,\left[E_{\rm J,0}\left(1 -
\frac{T_1^2}{T_0^2}\mathbf{S}^2\right) + \delta E_{\rm
J}(S_z^2-\mathbf{S}^2)\right]\sin\phi(t)\ ,
\end{equation}
where $\delta E_{\rm J}\equiv -T_1^2 B^2 \int dt \beta^{R}(t)\,t^2
=(\pi^2/16)\,T_1^2\rho^2\,(B^2/\Delta)$. Here, we clearly
elucidated the manner in which the spin dynamics alters
the Josephson current.

For $S=1/2$ the semi-classical approximation is insufficient. In
this case it is easier to perform a calculation with spin
operators~\cite{Bulaevskii}, rather than a path integral. One,
then, obtains~\cite{Bulaevskii} an expression for the Josephson
current identical to Eq.~(\ref{EQ:Josephson_Current}) with
$\mathbf{S}(t)$ being, however, the spin operator in the
interaction representation. Using the commutation relations of the
spin operators one obtains an extra contribution to the Josephson
current proportional to $S_z$. This allows reading out of the spin
state via the Josephson current. This extra contribution scales as
$S$ while the spin dependent contributions in
Eq.~(\ref{EQ:JC_corrections}) scale as $S^2$.

\section{detection}

We now briefly discuss a detection scheme for the
Josephson nutations for $S \gg 1$, 
e.g., in the semi-classical limit. In principle the 
nutations should affect the Josephson current.
The level of approximation employed in this paper 
was, however, insufficient to describe this effect.
Indeed, one has to substitute 
$\mathbf{S(t)}$ containing the nutations 
into Eq.~(\ref{EQ:Josephson_Current}). As the amplitude of the 
nutations is of the order $g_1$, the 
correction to the current will be of the order $g_1^2$.
We will study this correction elsewhere. 

Here we discuss a more direct detection strategy.
The spin motion generates a time-dependent magnetic field, $\delta{\bf B}({\bf 
r},t) = \frac{\mu_{0}}{4 \pi} [3 {\bf r} ({\bf r} \cdot {\bf m}(t))
-   r^{2} {\bf m}(t)]/r^{5}$, superimposed on
the constant external field ${\bf B}$.
Here, ${\bf r}$ is the position relative to the spin, with magnetic moment
${\bf m}(t) = \mu {\bf S}(t)$.
A ferromagnetic cluster of spin $S = 100$ generates a detectable field 
$\delta B \sim 10^{-10}$~T appears a micron
away from the spin. For a SQUID loop of micron dimensions located 
at that position, the corresponding flux variation 
$\delta \Phi \sim 10^{-7} \Phi_{0}$ (with
$\Phi_{0}$ a flux quantum) are within reach of modern SQUID's. For
such a setup with $T_{1}/T_{0} \sim 0.1$, the typical critical
Josephson current
is $J_{S}^{(0)} \sim 10 \;\mu\mbox{A}$,
$\vert\Delta\vert=1\;\mbox{meV}$, and $eV \sim 10^{-3}
|\Delta|$, we find that $\lambda S \sim 0.1$. Since $S_{x} =  S \sin \theta
\cos \phi$, $S_{y}=  S \sin \theta \sin\phi$, the spin
components orthogonal to ${\bf B}$ vary, to first order in
$(\lambda S)$, with Fourier components at frequencies $|\omega_{L} \pm
\omega_{J}|$ ($\omega_L=B$), leading to a discernible signal in the magnetic
field ${\bf B} + \delta {\bf B}$.  For a field $B \sim 200$~G, $\omega_{L} \sim 
560\; \mbox{MHz}$, and a new side band  will
appear at $|\omega_{L} - \omega_{J}|$, whose magnitude may be tuned
to 10--100~MHz. This measurable frequency is
markedly different from the Larmor frequency $\omega_{L}$.

\begin{figure}
\centerline{\includegraphics[width=0.5\columnwidth]{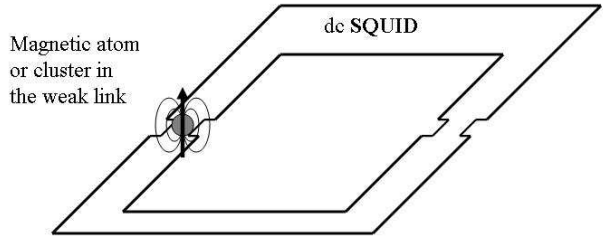}}
\caption{A SQUID-based detection scheme. The SQUID monitors the magnetic field
produced by the magnetic cluster in one of the junctions.}
\label{FIG:DETECTION}
\end{figure}

The efficiency of the detector may be further improved by  
embedding the spin in one of the Josephson junctions of the 
SQUID itself. The setup is sketched in Fig.~\ref{FIG:DETECTION}.
The Josephson junction containing the spin is used 
both for driving the nutations and, together with the second 
junction of the SQUID, for detecting them. 

\section{Discussion}

In this article, we illustrated that the dynamics of a spin embedded in a
Josephson junction is richer than appreciated hitherto.  We reported unusual
non-planar spin motion (in a static field), which might be probed directly and
which was further shown to influence the current in the Josephson junction.
Using a path-integral formalism, we described this non-planar spin dynamics and
the ensuing current variations that it triggers.  To describe the time evolution
we derived the effective action for a spin of arbitrary amplitude $S$ on the
Keldysh contour.  In passing, we noted a similarity between the resultant
effective action and that encountered in quantum antiferromagnetic spin chains.
Our central results are encapsulated in the effective action (\ref{EQ:S_beta}).

In the semi-classical limit of large $S$, relevant to
ferromagnetic spin clusters~\cite{Zhu_Nutation}, we obtained two
coupled equations of motion (Eqs.~(\ref{Eq:Eq_motion_spin}) and
(\ref{Eq:Eq_motion_phase})). These equations may be solved
perturbatively, as outlined above, or numerically. We presented an exact
limiting-case solution and illustrated how the new spin dynamics
may be experimentally probed.

The formalism developed can also be applied to the minimal $S=1/2$
system. In this case, however, it is simpler to perform a
calculation with spin operators \cite{Bulaevskii}, rather than a
path integral.

{\bf Acknowledgments}: This work was supported by the US Department of
Energy under LDRD X1WX, the CFN of the DFG, and the S.~Kovalevskaya award (YM). 
We thank L.N.~Bulaevskii and G.~Sch\"on for discussions.

\bibliography{spin}

\end{document}